\documentclass[12pt]{article}
\usepackage{sc3conf}
\usepackage{amsfonts}
\usepackage{epsfig}

\newcommand\GAIA{{\it GAIA}}

\newcommand\msunpctwo{{M_\odot{\rm pc}^{-2}}}

\newcommand\muas{\;\mu{\rm as}}
\newcommand\msun{{M_\odot}}
\newcommand\fwyn{f_{II}}

\begin{document}
\raggedbottom

\title{Five Years at the Movies}

\authors{N.W.~Evans \adref{1,2}
  and V.~Belokurov\adref{2}}

\addresses{\1ad Institute of Astronomy, Madingley Rd, Cambridge CB3 0HA
  \nextaddress \2ad Theoretical Physics, 1 Keble, Oxford OX1 3NP.}

\maketitle

\begin{abstract}
Applications of the forthcoming \GAIA\ satellite to the study of
variable phenomena (microlensing and supernovae) are discussed.
\end{abstract}

\section{Introduction}

\GAIA\ is an astrometric satellite which has recently been approved by
the European Space Agency for launch in about 2010. It will measure
the angles between objects in fields that are separated on the sky by
about a radian. Data will stream continuously at 1~Mbps from \GAIA's
three telescopes, providing information on the positions of the
billion or more astrophysical objects brighter than 20th
magnitude. The motion of objects across the sky, due to both their
space motion and their parallactic motion and the variability in the
brightnesses of objects in 15 wavebands will be measured because each
object is observed at least 150 times during the 5 year mission
lifetime. From the raw time series, a three-dimensional movie of the
motions of stars in the Galaxy will be synthesized. Just as the human
brain interprets signals from the retina in terms of moving
three-dimensional objects, so \GAIA\ will interpret signals from its
sensors in terms of moving astrophysical objects. \GAIA's job is
dramatically the more difficult because (1) \GAIA's sensors return
time-series rather than images, (2) the global adjustment of the
myriad of instantaneous positions is a computational task of huge
complexity, (3) \GAIA\ has to correct for gravity-induced distortions
of space-time, and (4) \GAIA\ sees in 15 colours rather than three.

This article presents highlights from the five-year \GAIA\ movie show.
The alert despatcher in the \GAIA\ mission will provide forewarning of
many kinds of bursting and variable phenomena. We discuss two
applications in detail here. First, the astrometric microlensing
signal will allow us to take a complete inventory of all objects -- no
matter how dark -- in the solar neighbourhood.  Second, the catalogue
of supernova (SN) detections will be the largest ever taken. It will
provide opportunities both for follow-ups in other wavebands and
numerous examples of scarce phenomena (like SNe II-L or SNe Ib/c).

\section{Dark Matter in the Solar Neighbourhood}

There are many direct detection experiments searching for the particle
dark matter in the solar neighbourhood~\cite{1}. A crucial question
is: how much particle dark matter is there?  This is not an easy
question to answer. Within 1.1 kpc of the Galactic plane, the column
density of all matter is $\sim 71 \msunpctwo$, of which $\sim 41
\msunpctwo$ is already accounted for by the known populations of stars
and gas~\cite{2}. This leaves $\sim 30 \msunpctwo$ as the column
density of dark matter. {\it How much of the missing matter is made up
from elementary particles, as opposed to dark compact objects like dim
stars?}

\GAIA\ is the first instrument which has the capability of answering
this question. Local populations of black holes and halo or disk white
dwarfs could easily have eluded identification thus far. Similarly,
there may be an extensive local population of very cool brown
dwarfs. The {\it 2 Micron All Sky Survey} and the {\it Sloan Digital
Sky Survey} have increased the known population of local dwarf stars
with spectral types later than M to over a
hundred~\cite{3}. Nonetheless, the coolest brown dwarfs will have
easily eluded the grasp of these and all current surveys. Future
projects like the {\it Space Infrared Telescope Facility} may be able
to detect very old brown dwarfs in the solar neighbourhood by using
long integration times, but a large-scale survey will be prohibitively
costly in terms of time.  However, the astrometric microlensing signal
seen by \GAIA\ will be sensitive to local populations of even the
dimmest of stars and darkest of these objects.

\begin{figure}[tb]
  \begin{center}
    \epsfig{file=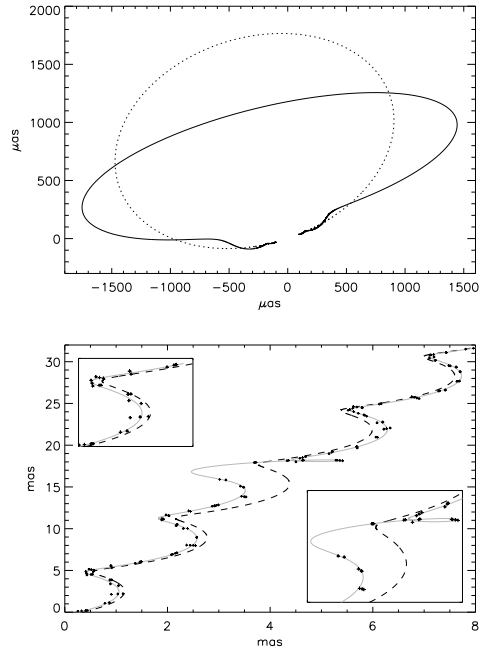, height=9cm}
\caption{Upper panel: Astrometric shift of a microlensing event, as
seen by a barycentric (dotted line) and a terrestrial observer (solid
line).  Lower panel: Simulated data incorporating typical sampling and
astrometric errors for \GAIA. Also shown for comparison are the
theoretical trajectories of the source with (grey line) and without
(dashed line) the event. The insets show the deviations at the
beginning and the midpoint of this high signal-to-noise event. (The
accuracy of the astrometry is $300\muas$, corresponding roughly to a
$17$th magnitude star).}
\end{center}
\end{figure}

\GAIA\ cannot resolve the two images of a microlensed source. However,
\GAIA\ can measure the small deviation (of the order of a fraction of
a milliarcsec) of the centroid of the two images around the trajectory
of the source. Astrometric microlensing is the name given to this
excursion of the image centroid~\cite{4}.  The cross section of a lens
is proportional to the area it sweeps out on the sky, and so to the
product of lens proper motion and angular Einstein radius. Each of
these falls off with lens distance, so the signal is dominated by
close lenses (within $\sim 1$ kpc of the Sun).  The all-sky
source-averaged astrometric microlensing optical depth is $\sim
10^{-5}$, over an order of magnitude greater than the photometric
microlensing optical depth~\cite{5}.  Fig.~1 shows two views of an
astrometric microlensing event. The upper panel shows the right
ascension and declination recorded by a barycentric and a terrestrial
observer (or equivalently a satellite at the ${\rm L}_2$ Lagrange
point, like \GAIA). However, \GAIA\ does not provide the data in such
a clean form.  The observed quantity is the CCD transit time for the
coordinate along the scan. (This is the same way that the {\it
Hipparcos} satellite worked~\cite{6}). From the sequence of these
one-dimensional measurements, the astrometric path of the source,
together with any additional deflection caused by microlensing, must
be recovered.  The lower panel shows the event as seen by \GAIA. The
simulated datapoints have been produced by generating random transit
angles, and sampling the astrometric curve according to \GAIA's
scanning law. The transits are strongly clustered, as \GAIA\ spins on
its axis once every 3 hours and so may scan the same patch of sky four
or five times a day. The transit angle is the same for all transits in
such a cluster, but changes randomly from cluster to cluster. The two
insets show the astrometric deviations at the beginning and at the
maximum of the event, from which it is clear that \GAIA\ can detect
that a microlensing event has occurred.

\begin{figure}
\epsfxsize=8.5cm \centerline{\epsfbox{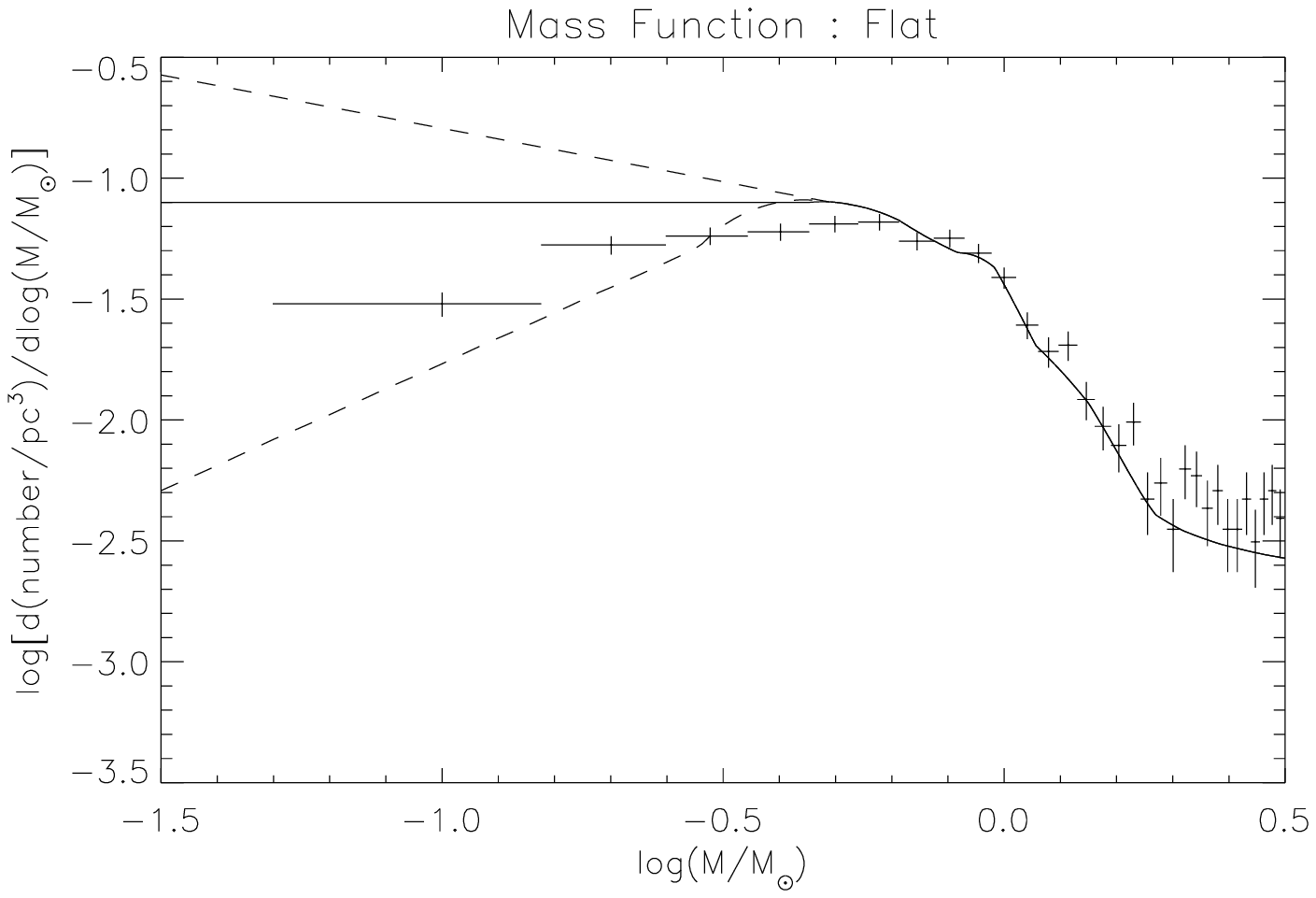}}
\epsfxsize=8.5cm \centerline{\epsfbox{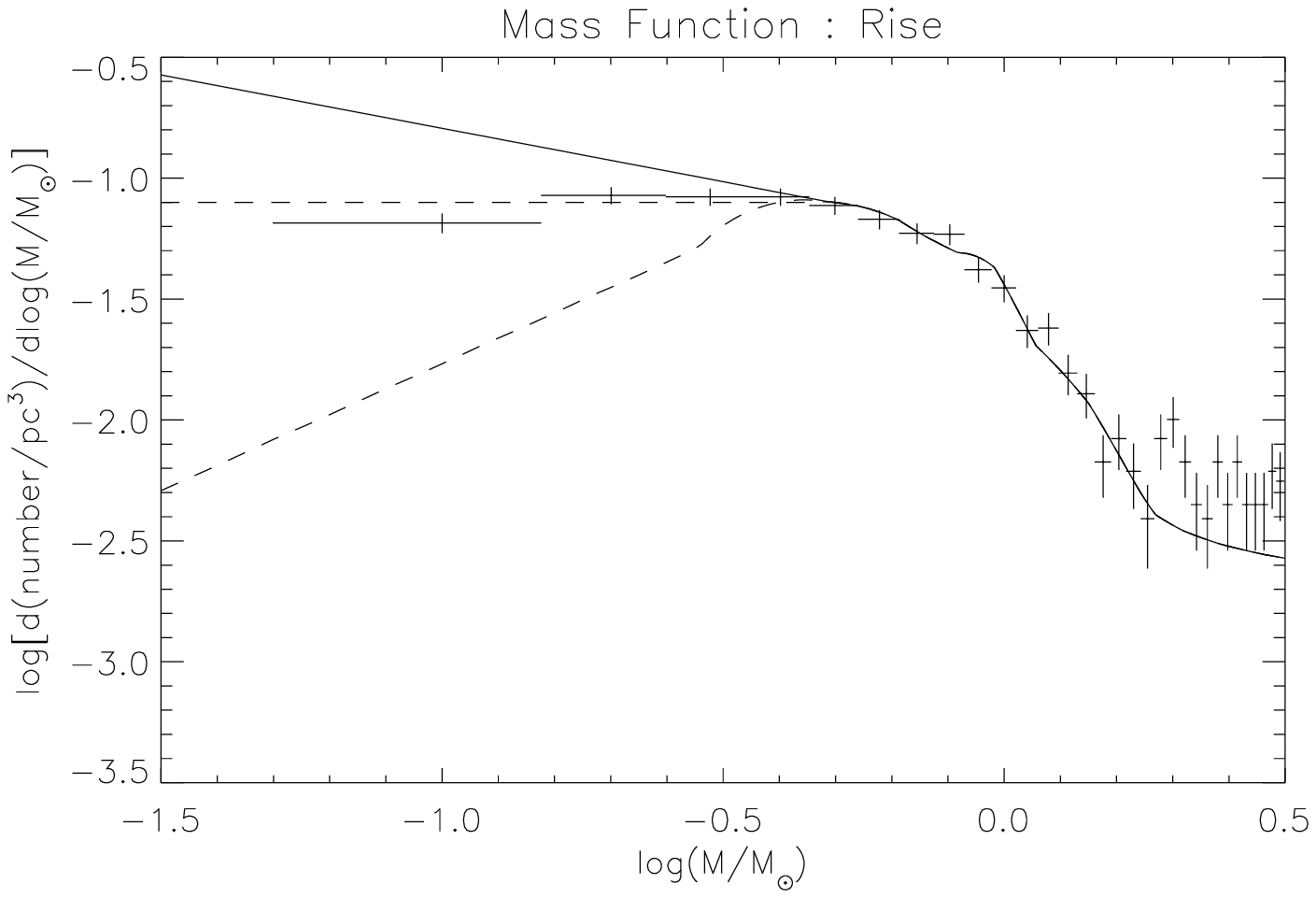}}
\epsfxsize=8.5cm \centerline{\epsfbox{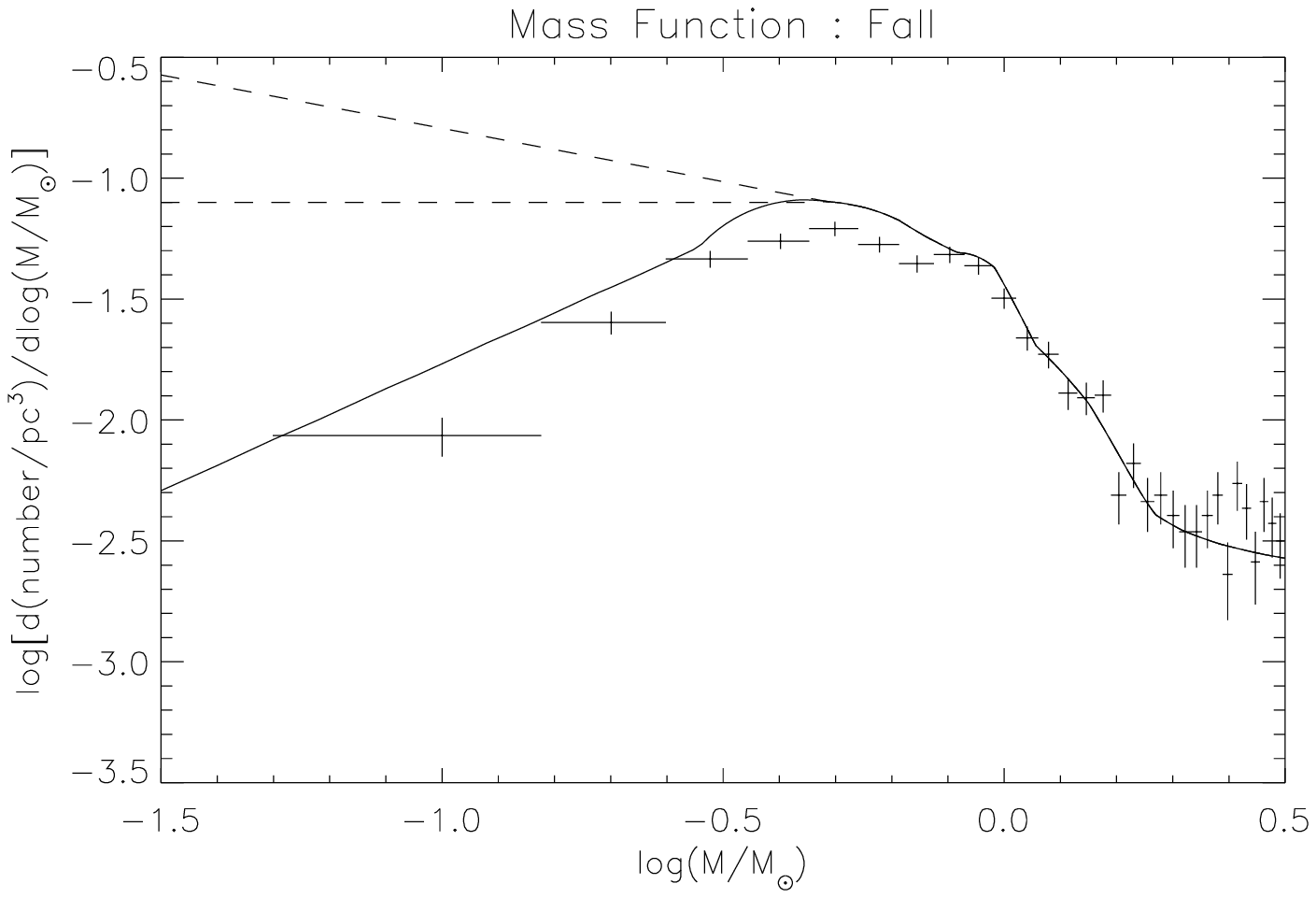}}
\caption{The recovery of the flat, rising and falling mass functions
from the subsamples of high quality astrometric microlensing events
generated from simulations.  The vertical error bar associated with
each bin is proportional to the square root of events in the bin.  The
horizontal error bar is the size of the bin.}
\label{fig:massfuns}
\end{figure}

Can the local mass function be recovered from the astrometric
microlensing signal seen by \GAIA? A covariance error analysis shows
that $\sim 2500$ astrometric events detected by \GAIA\ will be of such
high quality that the mass of the lens will be recovered with good
accuracy~\cite{5}. These events are biased towards lenses with mass
greater than $0.3 \msun$. This is not too surprising, as it is more
difficult to measure the deviation if the Einstein radius is small.
This already suggests that \GAIA's astrometric microlensing signal
will be an unbiased and powerful probe of the local population of
stellar remnants, such as white dwarfs and neutron stars. However, the
signal seen on smaller mass-scales will required modelling and
correction for selection effects before it is unbiased.  Reid \&
Hawley~\cite{6} compute the mass function (MF) for nearby stars from
the volume-limited {\it 8 Parsec Sample} using empirical
mass-luminosity relationships. They obtain a power-law MF ($f(M)
\propto M^{-1}$) over the range $0.1$ to $1.0 \msun$. This is in rough
agreement with the MF as inferred from studies of the luminosity
function (LF) of red disk stars seen with {\it Hubble Space
Telescope}~\cite{7}. The behaviour of the MF becomes more uncertain as
the hydrogen-burning limit is approached.  So, we use three MFs that
span the range of reasonable possibilities. Above $0.5 \msun$, the MF
is always derived from the Reid-Hawley luminosity function. Below $0.5
\msun$, it may be flat ($f(M) \propto M^{-1}$), rising ($f(M)\propto
M^{-1.44}$) or falling ($f(M) \propto M^{0.05}$). For the falling MF,
we choose the power-law index to be that implied by the decline in the
LF below $M_v \approx 13$. For the rising MF, we choose the power-law
index by assuming that the rise at $M_v \approx 11$ continues to
fainter magnitudes and that these stars are missing from the data
because of incompleteness. Of course, the flat MF is so called because
it has equal numbers of objects in each decade of mass and so appears
flat in a plot of $\log Mf(M)$ versus $\log M$.  For each of the three
MFs, we generate samples of 2500 astrometric microlensing events from
models of the Galaxy and compute the mass uncertainty using the
covariance analysis. (In actual practice, the high quality events
would be selected on the basis of the goodness of their $\chi^2$
fits).  We build up the MF as a histogram.  The vertical error bar
associated with each bin is proportional to the square root of events
in the bin.  The horizontal error bar is the size of the bin.

The three cases are shown in Fig. 2 with the solid line representing
the underlying MF. The simulated datapoints with error bars show the
MFs reconstructed from the high quality events.  It is evident that
\GAIA\ can easily distinguish between the flat, rising and falling
MFs. The MFs are reproduced accurately above $\sim 0.3 \msun$. Below
this value, the reconstructed MFs fall below the true curves, as a
consequence of the bias against smaller Einstein radii. However, this
does not compromise \GAIA's ability to discriminate between the three
possibilities. In practice, of course, simulations could be used to
re-calibrate the derived MFs at low masses and correct for the
bias. We have also carried out simulations with MFs containing spikes
of compact objects, such as populations of $\sim 0.5 \msun$ white
dwarfs.  They lie in the mass r\'egime to which \GAIA's astrometric
microlensing signal is most sensitive. So, such spikes stand out very
clearly in the reconstructed MFs.  We conclude that one of the major
scientific contributions that \GAIA\ can make is to determine the local
MF.  Microlensing provides the only way of measuring the masses of
individual objects irrespective of their luminosity.  \GAIA\ is quite
simply the best survey instrument to carry out an inventory of masses
in the solar neighbourhood.

\section{Supernovae Detection}

Supernovae (SNe) are divided into type I and II primarily on the basis
of their spectra, with lightcurve shape as a secondary diagnostic.
SNe Ia are thermonuclear explosions in white dwarf stars which have
accreted too much matter from a companion. It is unclear whether the
companion is also a white dwarf or is a main sequence/red giant
star. In fact, whether the progenitors of SNe Ia are doubly degenerate
or singly degenerate binaries is one of the major unsolved problems in
the subject.  SNe Ib/c and II originate in the core collapse of
massive stars. SNe Ia have found ready application in cosmology. Their
intrinsic brightness means that they can be detected to enormous
distances.  Although their peak luminosities vary by a factor of 10,
Phillips~\cite{9} found a correlation between the peak absolute
magnitude and the rate of decline, which enables SNe Ia to be
calibrated and used as ``standard candles''.  Claims for an
accelerating universe partly rest on their use as distance estimators,
yet there are obvious concerns regarding systematic
errors~\cite{10}. So, a major thrust of modern SNe studies is to
understand and quantify the differences in the morphology of SNe Ia
lightcurves and the scatter in the Phillips relation.  \GAIA\ is an
ideal tool to study nearby SNe (within a redshift $z \sim 0.14$).
\GAIA\ will provide a huge dataset of high quality local SNe Ia in
which any deviations from ``standard candles'' can be analysed.  Here,
we ask the question: {\it How many supernovae will \GAIA\ detect over
the five-year mission lifetime?}

Richardson et al.~\cite{11} use the Asiago Supernova Catalogue to
study the V band absolute magnitude distributions according to type.
They find that the mean absolute magnitude of SNe Ia at maximum is
$-18.99$. This figure includes the contribution from the internal
absorption of the host galaxy.  As \GAIA's limiting magnitude is $V
\sim 20$, this means that the most distant SNe Ia accessible to \GAIA\
are $\sim 630$ Mpc away. Similarly, the mean absolute magnitude of SNe
Ib/c at maximum is $-17.75$, so that the most distant SNe Ib/c
detectable by \GAIA\ are $\sim 355$ Mpc away. SNe II are subdivided
further according to lightcurve into linear (L) and plateau (P) types.
The mean absolute magnitude of II-L type is $-17.63$ and of II-P type
is $-16.44$ (Richardson et al. 2002).  These correspond to distances
of $\sim 335$ Mpc and $\sim 195$ Mpc respectively.  Such distances
emphasise that \GAIA\ is the ideal tool to discover relatively nearby
SNe.

The rates with which SNe occur in different galaxies at low redshift
are given in van den Bergh \& Tammann~\cite{12}.  The host galaxies
follow the local large-scale structure.  We use the latest version of
the CfA redshift catalogue~\cite{13}. This contains the sky positions
and heliocentric velocities of $\sim 20\,000$ galaxies. On plotting
numbers of galaxies versus distance, the graph peaks at $\sim 75$ Mpc
and thence shows a steady decline. This suggests that the CfA
catalogue can be used out to at most $\sim 75$ Mpc.  The numbers in
the catalogue must be regarded as lower limits to the true numbers
within 75 Mpc, as the luminosity functions (LFs) derived from the
catalogue are incomplete at faint magnitudes. Beyond 75 Mpc, we assume
that the distribution of galaxies is homogeneous and that the number
scales like $D^3$ where $D$ is the heliocentric distance.  Given the
numbers of galaxies and the observed rates, we can then
straightforwardly compute the total number of SNe that explode during
the five-year \GAIA\ mission lifetime and are brighter than the
limiting magnitude.  In all, there are at least $\sim 48\,000$ SNe Ia
and $\sim 7000$ SNe Ib/c. The numbers for SNe II depend on the
relative frequency of II-L with respect to II-P, which is not very
well-known. Henceforth, we denote the fraction of all SNe II that are
L-type by $\fwyn$. Then, the numbers of SNe II that explode are $\sim
28\,500 \fwyn + 5600 (1-\fwyn)$.  These numbers are lower limits for
two reasons -- first because no contribution from faint galaxies is
included and second because \GAIA's limiting magnitude may be deeper
than 20th in practice.

\begin{figure*}
\begin{center}
\epsfxsize=12cm \centerline{\epsfbox{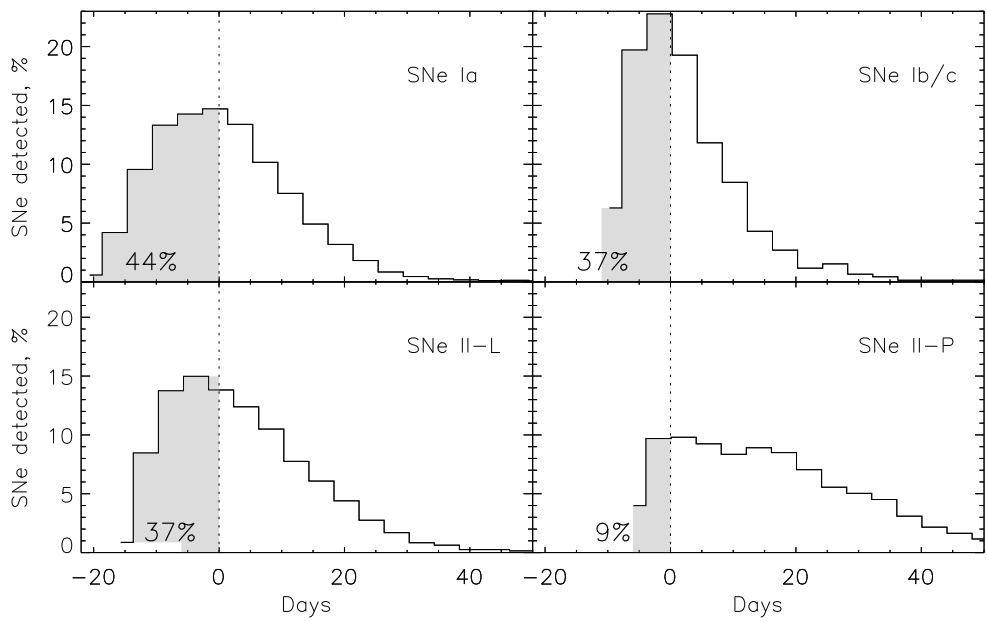}}
\end{center}
\caption{This shows histograms of the numbers of detected SNe against
phase of the lightcurve. The shaded area corresponds to the fraction
of SNe caught before the maximum of the lightcurve.}
\label{fig:phase}
\end{figure*}

By using simulations of SNe detection with \GAIA's scanning
law~\cite{14}, we find that \GAIA\ records data on $30 \%$ of all the
SNe Ia within 630 Mpc, which marks the limit of the most distant SNe
Ia accessible.  Similarly, \GAIA\ records data on $\sim 20 \%$ of all
SNe Ib/c, $\sim 31 \%$ of all SNe II-L and $\sim 48 \%$ SNe II-P
within these distances.  This means that \GAIA\ will provide some
(perhaps rather limited) information on $14\,300$ SNe Ia and $1400$
SNe Ib/c during its five-year mission.  For SNe II, the number depends
on the relative frequency $\fwyn$ and is $\sim 8700 \fwyn + 2700
(1-\fwyn)$.  If SNe II-L and SNe II-P occur equally frequently ($\fwyn
= 0.5$), then the total number of SNe II is $\sim 5700$. In other
words, \GAIA\ will provide some information on $\sim 21\,400$ SNe in
total.  These are huge numbers, both compared to the sizes of existing
catalogues and to the likely datasets gathered by other planned space
missions.  Almost all the SNe that \GAIA\ misses explode in the 20
days just after \GAIA\ samples that location in the sky. Before the
next transit of the telescopes, the SN reaches maximum and then fades
to below \GAIA's limiting magnitude ($V \sim 20$).  

Fig. 3 shows the fraction of the detected SNe as a function of phase
of the lightcurve. Some $44 \%$ of the detected SNe Ia are caught
before maximum, $37 \%$ of the detected SNe Ib/c, $37 \%$ of the
detected SNe II-L and $9 \%$ of the detected SNe II-P. The low
fraction for SNe II-L is largely a consequence of the fact that they
are intrinsically the faintest. The total number of all SNe found
before maximum during the 5 year mission lifetime is $\sim
8\,500$. This number can be broken down into $\sim 6\,300$ SNe Ia,
$\sim 500$ SNe Ib/c and 1700 SNe II (assuming $\fwyn = 0.5$).

\GAIA\ will provide a large dataset of nearby SNe, which are very
interesting from the point of view of understanding the properties and
the underlying physics of the explosions themselves.  The advantage of
SNe surveys with \GAIA\ is that selection effects are either minimised
or easy to model, and that there will be many examples of
comparatively scarce phenomena (e.g., subluminous SNe, SNe II-L, SNe
Ib/c).  At present, SNe rates come from relatively small datasets and
are subject to substantial uncertainties. Selection effects --
depending on the type of host galaxy, the extinction and the distance
from the center of the galaxy -- seriously afflict all current
datasets.  Given the large numbers of alerted SNe, \GAIA\ will provide
accurate rates as a function of position, extinction and type of host
galaxy.  These give valuable, if indirect, information on both the
star formation rate and the high mass end of the mass function. There
have also been suggestions that populations of subluminous SNe may
have been systematically missed in existing catalogues~\cite{11}. If
so, then \GAIA\ is the ideal instrument with which to find them.

\section{Conclusions}

The \GAIA\ satellite is scheduled for launch in about 2010. It will
determine the positions, velocities and astrophysical nature of over a
billion stars distributed throughout the Milky Way Galaxy and into the
Local Group. Each object will be monitored between 150 and 400 times
during the 5 year mission (depending on ecliptic latitude). The
resulting catalogue will be one of the largest astrophysical datasets
ever taken. The positions and velocities of stars, together with the
changing brightness of variable sources (e.g., bursting and eruptive
stars, supernovae, killer asteroids, eclipsing variables, microlensed
stars), will all be synthesized into a three-dimensional movie show.

This article has discussed two applications of real-time detection of
variable sources with the \GAIA\ satellite. \GAIA\ will provide the
best determination of the local mass function of stars down to brown
dwarf mass scales -- which is important for reckoning the baryonic and
particle dark matter content of the solar neighbourhood. This can be
inferred from the astrometric microlensing signal of nearby stars.
\GAIA\ will provide the largest sample of nearby supernovae ever
acquired -- which is crucial for assessing how the scatter in
properties affects their use as cosmological distance indicators.
This will provide opportunities both for follow-ups in other wavebands
(e.g., X ray and gamma-ray) and with other detectors (e.g., neutrino,
gravitational waves).  Further applications are given in
references~\cite{5} and~\cite{14}.

\end{document}